# Crowdfunding as Entrepreneurial Investment: The Role of Local Knowledge Spillover

Filippo Marchesani[(✉)] and Francesca Masciarelli

University of "G. d'Annunzio" Chieti – Pescara, Viale Pindaro, 42, 65127 Pescara, Italy
`{filippo.marchesani,francesca.masciarelli}@unich.it`

**Abstract.** This paper explores the role of local knowledge spillover and human capital as a driver of crowdfunding investment. The role of territory has already been studied in terms of campaign success, but the impact of territory on the use of financial sources like equity crowdfunding is not yet known. Using a sample of 435 equity crowdfunding campaigns in 20 Italian regions during a 4-year period (from 2016 to 2019), this paper evaluates the impact of human capital flow on the adoption of crowdfunding campaigns. Our results show that inbound knowledge in the region, measured in terms of ability to attract national and international students, has a significant effect on the adoption of crowdfunding campaigns in the region itself.

## 1 Introduction

Crowdfunding as a method of entrepreneurial financing is growing very quickly following the high development of internet tools in the financial market [1]. Entrepreneurs are using Internet platforms to appeal to the "crowd" by listing and describing their investment or cause. This way, they can reach a large audience where each individual provides a portion of the requested amount to fund their start-up using crowdfunding as an alternative to a traditional form of new ventures' financing, i.e., venture capital financing [2].

Crowdfunding platforms have become diverse and specialized, increasingly targeting differentiated segments which cover several different forms such as donation-based crowdfunding, rewards-based crowdfunding, debt-based crowdfunding, and equity-based crowdfunding. Our focus is on equity-based crowdfunding which represents the main form of crowdfunding campaigns in terms of capital and financing. Based on the definition of Ahlers [3, p. 955]: "Equity crowdfunding is a form of financing in which entrepreneurs make an open call to sell a specified amount of equity or bond-like shares in a company on the Internet, hoping to attract a large group of investors", we evaluate the influence of local knowledge on entrepreneurial investment that has completely modified the original relationship between territory and new ventures.

Until a few years ago, the relationship between local banks and start-ups was very steady because a large part of entrepreneurs preferred to borrow money from their local banks [4]. A stream of the literature shows that geographical distance matters to small business lending, although technology weakens the dependence of small businesses on local lenders [5, 6].

Therefore, when entrepreneurs have to incur higher transaction costs to borrow from local lenders, they may search for alternative sources of financing such as crowdfunding [7]. We already know that crowdfunding serves as a viable alternative to traditional sources and, in terms of local development, that a large part of crowdfunding activities are present in regions that have more concentrated credit markets [8]. Furthermore, it is important to evaluate the relationship between crowdfunding campaigns and the region to explain which characteristics in terms of knowledge and human capital promote the use of innovative financial instruments such as crowdfunding.

The relationship between crowdfunding campaigns and start-up location has been studied in recent years [9, 10]. We know that the territoriality of a campaign influences its success because local investors invest relatively early, and they appear less responsive to the decisions of other investors [10], but we are not well aware of what encourages the development of a crowdfunding campaign in certain regions and what leads an entrepreneur to use creative finance tools in these regions rather than turning to the financial market.

There are only a few studies that have tried to explain this relationship. For example, Sorenson et al. 2016 [11] states that crowdfunding appears to be relatively stronger in regions with less venture capital funding, compared to traditional hubs such as Silicon Valley and Boston. Our paper helps fill this gap by evaluating knowledge spillover in Italian regions to answer these overarching research questions: "*Do the characteristics of the geographical area in terms of knowledge attraction affect the adoption of crowdfunding?*" and "*Are regions with high foreign student flows more likely to adopt alternative financing sources such as crowdfunding campaigns?*".

To test our conjectures, we considered the most widely used Italian equity crowdfunding platforms, namely Mamacrowd and CrowdfundMe. This study was conducted using the ordinary least squares method (OLS) to verify the research framework and hypotheses. We have analyzed 20 Italian regions for 4 years (from 2016 to 2019). Hypothesis testing was conducted using a panel data regression analysis model that aims to predict the extent of the strength of knowledge spillover effects on the use of uncommon financial sources such as equity crowdfunding.

Our results show that inbound knowledge measured in terms of student mobility from other regions has a significant effect on the use of crowdfunding campaigns in the region itself. Moreover, we evaluated the impact of transnational students in the region. Our results show that both variables, inbound students from other regions and other countries, have an impact on the regional development in terms of access to the crowdfunding market. Regions with a higher number of students from other regions and foreign students are more likely to use creative finance tools, such as crowdfunding, than regions with fewer incoming student flows.

By exploring this relationship, we believe that our study is capable of offering new interesting theoretical insights to the nascent literature on entrepreneurial finance. In addition, our findings have several implications for regional development and innovation policies. Regions with a large part of inbound knowledge flow are more likely to support the innovation process and the use of alternative financial sources, such as crowdfunding [12], which elements contribute to the development of the crowdfunding campaign in a certain area.

## 2   Theory and Hypothesis

The global crisis began in 2008 and drastically reduced bank borrowing [13] and venture capital investments [14], thus prompting entrepreneurs to seek alternative sources of finance for their start-ups. This has fueled the rise of equity crowdfunding, a crowdfunding model in which entrepreneurs make an open call to sell equity shares of their start-ups to the crowd of Internet users [3].

Following the definition of Belleflamme [7], crowdfunding "involves an open call, mostly through the Internet, for the provision of financial resources either in the form of donation or in exchange for the future product or some form of reward to support initiatives for specific purposes" (2014, p. 588). From this definition we understand that the "structural" factor is represented by the technological innovation of Web 2.0, which has allowed internet users to generate online content and share it with other users, de facto making crowdfunding viable [15, 16].

Research examining entrepreneurial fundraising efforts, including crowdfunding, has frequently drawn on different theoretical lens to understand investment transactions between investors and entrepreneurs [17–19].

In the last decade, equity crowdfunding has become an established source of funding for entrepreneurial firms [20], giving individual investors the opportunity to buy shares of unlisted companies online. As a result, the crowdfunding market has grown fast in recent years. To date, there are approximately 2000 crowdfunding websites, and the World Bank expects that crowdfunding could account for over $300bn in cumulative transactions by 2025 [21].

The entrepreneurial finance literature has highlighted different characteristics related to the crowdfunding campaign that can be resumed in 3 main categories: campaign characteristics, networks, and understandability of the company's concept and offering [22]. In the related literature, several studies try to understand the role of team composition [23, 24], entrepreneurial skills [25] and trait [26], but very few studies focus on the development of this phenomenon in terms of region and the role of knowledge acquisition in the region itself.

### 2.1   Knowledge Spillover, Human Capital, and Regional Development

The concept of human capital was fully developed in the 1960s with the emergence of human capital theory formalized by Schultz [27] and Becker [28, 29]. Human capital represents the combined intelligence, skills, experience, education, and expertise of organizational employees [30]. It constitutes a high-value corporate asset and is accepted as a prime factor in the intellectual capital framework. Following Goldin at. al 2016 [31] defining human capital as "the stock of habits, knowledge, social and personality attributes (including creativity) embodied in the ability to perform labor so as to produce economic value", we posit that human capital represents an important economic driver on regional development, and we understand that the role of knowledge and intellectual capital plays a key role in this relationship. Following the previous literature, we assume that a "higher" human capital is associated with higher capabilities and skills concerning various aspects of the entrepreneurial opportunity, such as defining and implementing a venture's strategy [22], identifying and exploiting business opportunities [32, 33], acquiring additional financial resources [34, 35].

The literature on urban systems [36] considers these knowledge processes to be primarily an interregional macro-phenomenon, in which the geographical mobility of physical capital in terms of firms [37] and human capital in terms of labor [38, 39], determines the revealed patterns of economic geography. The literature on agglomeration and clustering puts the accent on a large number of interactions between agents within a spatially constrained environment [40]. These interactions affect local development in terms of knowledge exchange [41], industrial clusters [42], R&D investments [43], and innovation [44].

To understand the role played by human capital in economic development, it is important to build on the new growth theories. One of the most important theories was proposed by Romer [45–47] who first formally examined the role of knowledge in growth from the perception of the public good aspect of knowledge spillovers, and this type of approach has been a starting point in the agglomeration literature. Moreover, the role played by knowledge externalities connected with human capital has also been supported by Lucas [48] who emphasizes the role of human capital accumulation related to technological change, schooling, and through learning-by-doing.

To connect human capital and regional development, an important role has been played by the university. There is an extensive literature examining the role of universities in relation to economic growth [49, 50]. As already noted, universities play a key role as producers of creative and high human capital that is embodied in their graduates and staff and that has an impact on regional economic growth [12, 51, 52]. Besides, several research studies have also argued that universities can be a contributor to growth as a source of knowledge spillovers [39, 53, 54]. Universities are assumed to be important sources of localized knowledge spillovers due to their explicit focus on knowledge generation and dissemination.

The impact of knowledge spillover on regional development and the crucial role of the university in this phenomenon has gained a great level of importance in recent years.

It is crucial to study cross-regional student mobility in terms of university enrollment of national and international students since the human capital of new generations is one of the most important areas for exploration and development in advanced economies. King & Ruiz-Gelices [55] pointed out that worldwide student migration is an increasingly important phenomenon that requires attention from both policy and research perspectives. In addition to that, Dustmann & Glitz [56] stated that education and migration are so closely linked that it is almost impossible to separate them from each other. This linkage represents an important growth prospect in terms of knowledge and regional innovation.

Besides, the university plays a very important role in the development policies of a region, in the ability to innovate and attract "higher" human capital, and contributes to the evaluation of economic policies supporting entrepreneurship [57].

Regarding the connection between local context and innovation, most of the literature is based on patents and R&D as indicators of innovation. For example, Jaffe et al. 2000 [58] analyzed the geographic location of patent citations in order to show that knowledge spillovers are geographically localized. Moreover, Maurseth and Verspagen [59] conducted similar research for knowledge spillovers across European regions and came to the same conclusion. By tracking patent citations, these studies focused on the exchange of explicit knowledge. As for R&D, Harabi [60] investigated the effectiveness of different channels of R&D spillovers at the intra-industry level. The study observed R&D activities, reverse engineering, publications, technical meetings, interpersonal communication, and patent disclosures as possible channels for knowledge spillovers. The aforementioned study suggests that a firm's investment in R&D is the most important channel for spillovers. In addition, Florida and Kenney [61, 62] used venture capital data to document the geographic patterns of high-tech entrepreneurship and the social structures of innovation set by a venture capitalist supporting such geographically organized social structures of innovation.

We assume that the relationship between knowledge spillover, human capital, and local context is associated with regional development and the innovation process of the region itself. Furthermore, "higher" human capital from other countries is usually attracted to innovative areas to which it can contribute and from which it can learn, effectively participating in a virtuous cycle where a strong innovation system attracts skills, which make it stronger and, in turn, more attractive to human capital [63]. As a result, this relationship affects regional entrepreneurial investment and access to new financial tools such as crowdfunding.

Human capital has a great impact on the investor's decision and is one of the first aspects investors pay attention to before funding a company, as it is a key factor in determining a firm's success [64, 65]. We also know that the influence exerted by human capital is greater for young businesses compared to old ones [66, 67]. Human capital is a multifaceted concept, largely related to the capabilities and skills of the entrepreneurial team, leading to the success of new ventures, and their investors being remunerated for the uncertainty related to their prospects [3], but it is also interesting to assess the impact of territory and knowledge contamination on the access to financial tools such as crowdfunding.

## 2.2  Equity Crowdfunding and Regional Entrepreneurial Investment

In recent years, crowdfunding as a method of entrepreneurial financing has grown very rapidly. Recently, entrepreneurs have been using Internet platforms to appeal to the "crowd"; by listing and describing their investment or cause, entrepreneurs can reach a large audience where each individual provides a portion of the requested amount to fund their start-up using crowdfunding as an alternative to traditional venture capital financing [2]. Crowdfunding platforms have become diverse and specialized, and they increasingly target differentiated segments covering diverse forms. Donation-based crowdfunding is used to collect charitable funding in support of causes and projects. In rewards-based crowdfunding, funders receive non-monetary rewards in exchange for their contribution. Debt-based crowdfunding offers a credit contract, while equity-based crowdfunding offers an equity stake in the target company. Our focus is on equity-based crowdfunding. Following Ahlers' [3, p. 955] previous definition we notice that the use of these financial tools has completely changed the original relationship between territory and start-up.

The role played by crowdfunding finance represents an alternative (or a complement) to more traditional funding sources such as debt finance and makes it possible to skip the common relationship between territory and entrepreneur. Although distant investors are common for publicly traded companies, theory predicts that investors in early-stage entrepreneurial ventures will tend to be local. That is because gathering information, monitoring progress, and providing input are particularly important to investors in early-stage ventures, and the costs of these activities are sensitive to distance, which encourages the entrepreneur to access financial tools such as crowdfunding. Most of the empirical evidence to date supports these claims [68–71]. But it is also interesting to note that in recent years this relationship has change as a result of the innovation process and the use of the Internet. This paper explores the role of knowledge spillover and human capital in the region as a driver of equity crowdfunding investment. The role of the territory has already been studied in terms of campaign success and access to the financial market, but the impact of the territory on the use of and access to financial sources such as equity crowdfunding is not yet known.

In general, capital access can be associated with the ability of the business to obtain an advantage from credit or loan offered by financial intermediaries [13]. Adequate access to external capital needs to be combined with elements internal to the business organization itself. As reminded by Green, Covin, and Slevin [72], entrepreneurial orientation may be a significant element in representing the architecture of firm management, but is also important to pay attention to external elements such as location and territory [73].

The role of territory in terms of geographical clusters has attracted much attention in the academic literature. A large number of studies have explained the reasons for regional and geographical cluster competitiveness and recognized that the very existence of the cluster is a distinctive feature of that region that positively affects its overall performance [74]. Specifically, geographical clustering has been show to encourage national, regional, and local competitiveness, innovation and growth and to sustain the competitive advantages of regions by fostering innovation [36, 75]. Prior studies have highlighted the significant effects of location advantages because regional networks prompt information flows [76], thus enabling spillover effects of start-up success and affecting the creation of new firms at the regional level [67, 77].

The relationship between territory and the use of alternative financial sources such as crowdfunding is quite new. Previous studies suggest that although crowdfunding mitigates geography-related friction [3], location influences crowdfunding outcomes: projects that were closer to banks attracted less funding from local investors [8], while those located where there is more creative population have a higher success rate [2]. Location effects are also manifested through local altruism or the promotion of projects that share similar values with local communities [78]. Recently, Giudici *et al.* 2018 [79] have shown that certain salient features of the geographical area where entrepreneurs reside influence the success of the crowdfunding projects they propose and have found that the existence of social relationships among people residing in a specific geographical area increases the likelihood of success of reward-based entrepreneurial projects. Moreover, the nascent crowdfunding literature has highlighted the existence of a self-reinforcing pattern whereby contributions received in the early days of a campaign accelerate its success [80].

We know that the territoriality of a campaign influences its success because local investors invest relatively early and, consequently, this responsiveness affects the success and development of a campaign [10, 19], but we are not well aware of what encourages the development of a crowdfunding campaign in certain regions and what leads an entrepreneur to use creative finance tools in these regions rather than turning to the financial market. The present paper helps fill this gap by evaluating the influence of knowledge spillover in Italian regions. Thus, we posit:

> HI: *The characteristics of the geographical area in terms of knowledge attraction positively affect the adoption of crowdfunding campaigns.*

Evidence exists that local characteristics affect entrepreneurs' ability to attract external financing [4, 81]. However, the crowdfunding literature is silent in this regard, even though it has shown that geography influences these Internet-based financing sources.

Besides, to understand the influence of knowledge spillover on the adoption of crowdfunding campaigns, we evaluate the impact of foreign students from other nations on each region in response to the following:

> HII: *Regions with a high number of foreign students' flows are more likely to adopt alternative financing sources such as crowdfunding campaigns.*

The region is a highly relevant support space for firms' innovation process because of the need to have continuous interaction to exchange knowledge and collaborate in joint innovation projects with people from other regions and other countries. In this case, the heterogeneity of students from other regions and other countries represents an important mechanism of knowledge transfer [82]. We assume that a large and heterogeneous flow of inbound knowledge and human capital would promote and support the innovation process and the possibility to turn to alternative creative financial sources such as crowdfunding.

## 3    Methodology

### 3.1    Sample and Data Collection

We analyzed 435 crowdfunding campaigns from the most popular equity-crowdfunding platforms in Italy such as Mamacrowd and CrowdfundMe, which represent the main crowdfunding platforms used in Italy. Mamacrowd was launched in 2011, while Crowd-fundMe was launched in 2013, and both platforms are authorized by Consob that plans to invest in innovative Italian projects such as start-ups and innovative SMEs ranking first and second, respectively, in the Italian equity crowdfunding market by transaction volume and number of campaigns. We examined the position of the proposing campaign and the development of the use of crowdfunding campaigns in the Italian regions.

Data were mainly collected from official ECF platform websites, but this research also relied on the integration of other data sources, such as reports, official business register databases, and National Student Clearinghouse (NSC) to explain the interaction between regions' knowledge flow and local crowdfunding development.

To recognize the role played by knowledge spillover and emphasize the importance of universities in regional development, student mobility was considered to assess the interconnection between regions. The National Student Clearinghouse database was used. NSC is an administrative archive where all the students enrolled in the Italian university system are registered. Data within the NSC database are submitted monthly by the Italian universities and all the students enrolled in a university during a given academic year, regardless of the year of the course, are taken into account. Our sample is based on more than 6,500,000 Italian students and focuses on 1,600,000 students who move from one region to another one and from another country to the region under consideration. They represent an uninterrupted channel for the transfer of scientific and technical knowledge among countries and regions [83, 84].

This study was conducted using the ordinary least squares method (OLS) to verify the research framework and hypotheses. We analyzed 20 Italian regions and crowdfunding campaigns during a 4-year period (2016 to 2019). Specifically, we analyzed data on campaigns listed on those two platforms from January 2016 to December 2019. Hypothesis testing was conducted using a panel data regression analysis model that aims to predict the extent of the strength of the effects of both independent variables on the dependent variable.

### 3.2  Operationalization of the Variables

The crowdfunding campaign is our unit of analysis. Our dependent variable is the campaign, and we have considered the total number of crowdfunding campaigns for each Italian region in the last 4 years. We examined 435 crowdfunding campaigns in the last 4 years. Specifically, 47 campaigns referring to 2016, 109 campaigns referring to 2017, 124 referring to 2018, and 155 referring to 2019 were analyzed. This trend represents an important growth in terms of the use of the crowdfunding campaign as a method of entrepreneurial financing in the last 4 years. This variable was measured by the logarithm of the ratio of crowdfunding campaigns in each region per the total number of enterprises in the respective region. This allowed us to analyze the number of campaigns based on the number of companies in the area.

Our model includes two independent variables: *students_regions* and *students_countries.* Both variables were measured by the interaction between each region and incoming student flow from other regions and other countries. These variables consider students enrolled in university courses at an Italian university. The sample analyzed ranges from 18 to 35 years. A total of 1,600,00 students were examined to understand the linkage between knowledge spillover and crowdfunding development within Italian regions.

Our first independent variable, students' region, is measured by incoming student flow from other regions in the region under consideration. As a proxy for inbound knowledge from other regions, we used the number of students from other regions over the population of the region itself. Knowledge spillover effects are significantly stronger when there is a critical mass of high technology [85] moderating the relationship between start-up and venture capital [86].

Our second independent variable, students' country, is measured by incoming student flow from another country to the region considered over the population of the region itself. They represent an uninterrupted channel for the transfer of scientific and technical knowledge among countries. In the most advanced countries, the number of foreign students enrolled in higher education has had a surprising growth rate over the last decade [86–88], influencing regional development in terms of innovation [90] and start-ups [91]. Both of our independent variables were measured by taking the log of the number of students over the population in the region.

In our regression model, we controlled for both region and campaign characteristics. More specifically, the model specification includes the following indicator as a control variable. Total firms considered the total number of firms registered in the Chamber of Commerce. Expenditure in R&D considered the expenditure of each region in R&D over the region's population. We also considered a Gross domestic product (GDP) of every region that represented the city under consideration. GDP is a monetary measure of the market value of all the final goods and services produced in a certain period. In addition, regarding the development of start-ups in each region, we considered New Firms with regard to the firm founded in the year taken into consideration over the total number of active firms over time based on the registration in the Chamber of Commerce.

Table 1 presents the summary statistics of the variables used in the regressions and Table 2 the correlation matrix related to this model.

**Table 1.** Descriptive statistics

| Variable | Obs. | Mean | Std. Dev. | Min | Max |
|---|---|---|---|---|---|
| Campaigns (log) | 80 | 0.001 | 0.002 | 0 | 0.006 |
| Students_Regions (log) | 80 | 0.263 | 0.144 | 0.154 | 0.592 |
| Students_Countries (log) | 80 | 0.147 | 0.013 | 0 | 0.043 |
| Total_Firms (log) | 80 | 0.095 | 0.077 | 0.073 | 0.767 |
| Expenditure_R&D (log) | 80 | 11.197 | 4.31 | 4.713 | 21.854 |
| New_Firms (log) | 80 | 0.061 | 0.004 | 0.054 | 0.072 |
| GDP (log) | 80 | 83545 | 84505 | 4359 | 383175 |

**Table 2.** Correlation model

|  | Variable | [1] | [2] | [3] | [4] | [5] | [6] | [7] |
|---|---|---|---|---|---|---|---|---|
| [1] | Campaigns (log) | 1 | | | | | | |
| [2] | Students_Regions (log) | 0.281 | 1 | | | | | |
| [3] | Students_Countries (log) | 0.274 | 0.354 | 1 | | | | |
| [4] | Total_Firms (log) | −0.056 | −0.136 | −0.084 | 1 | | | |
| [5] | Expenditure_R&D (log) | −0.001 | 0.228 | 0.431 | −0.068 | 1 | | |
| [6] | New_Firms (log) | 0.035 | −0.177 | −0.351 | −0.039 | −0.055 | 1 | |
| [7] | Gross Domestic Public (log) | 0.101 | 0.015 | 0.357 | 0.045 | 0.438 | 0.129 | 1 |

## 4   Results

The role of knowledge spillover in the relationship between crowdfunding and regional entrepreneurial investment has been explored in this paper. Our result shows that inbound knowledge measured in terms of student mobility from other regions has a significant effect on the use of crowdfunding campaigns in the region itself. Moreover, we evaluated the impact of transnational students in the region.

The empirical results are presented in Table 3. Table 2 reports the correlation between the variables, and the descriptive statistics are presented in Table 1. No major collinearity issues are detectable. To assess potential multicollinearity, we computed the variance inflation factors (VIFs). For each model (Table 3) the mean and maximum VIF are well below the threshold of 1,5. Therefore, we concluded that multicollinearity is not a threat to the validity of our results. Going beyond simple correlation, Table 3 reports the estimated results from regression.

**Table 3.** Regression matrix

|  | Model I | Model II | Model III | Model IV |
|---|---|---|---|---|
| Campaigns (log) | | | | |
| Students_Regions (log) | | 0.001** | | 0.001* |
| | | [0.001] | | [0.001] |
| Students_Countries (log) | | | 0.002*** | 0.001** |
| | | | [0.001] | [0.001] |
| Total_Firms (log) | | 0.002 | −0,001 | 0,002* |
| | | [0.001] | [0.001] | [0.001] |
| Expenditure_R&D (log) | −0.001 | −0.002 | 0.006 | −0.001* |

(*continued*)

**Table 3.** (*continued*)

|  | Model I | Model II | Model III | Model IV |
|---|---|---|---|---|
|  | [0,001] | [0,001] | [0,002] | [0,003] |
| New_Firms (log) | −0.001 | 0.002* | 0.001* | 0.001* |
|  | [0.001] | [0.001] | [0.001] | [0.001] |
| GDP (log) | 0.001 | 0.002 | 0.001 | 0.003 |
|  | [0.001] | [0,001] | [0,001] | [0.001] |
| No. of observation | 80 | 80 | 80 | 80 |
| R-squared | 0.221 | 0.352 | 0.328 | 0.471 |

Notes: Students DC, dependent variable. $P < 0.100$; * $P < 0.050$; ** $P < 0.010$; *** $P < 0.001$

Model I in Table 3 illustrates the result of a specification containing only the control variables considered. Model II shows the effect of our firs independent variable, that is, incoming students from other regions, and provides support for the hypothesis H1 (*The characteristics of the geographical area in terms of knowledge attraction positively affect the adoption of crowdfunding campaigns.*) by showing a positive and statistically significant association between inbound knowledge as measured by student mobility from other regions to the region under consideration and the use of alternative financial tools such as equity crowdfunding in the region itself ($p = 0.006$, β = 8.72). In other words, we found that a region with a large amount of incoming knowledge from other regions is more likely to use crowdfunding campaigns as a financial method compared to other regions.

Besides, Model III in Table 3 provides support for our hypothesis H2 (*Regions with a high number of foreign students' flows are more likely to adopt alternative financing sources such as crowdfunding campaigns*). The large presence of incoming students from other countries positively affects the adoption of equity crowdfunding campaigns in the region itself ($p = 0.001$, β = 6.24).

As we can see in Model IV, a high presence of incoming students from other regions and other countries influences the use of equity crowdfunding campaigns as alternative financing method and increases the possibility to access the crowdfunding market. Evidence exists that local characteristics affect entrepreneurs' ability to attract external financing [4, 81]. We found that entrepreneurs in regions with a larger number of students from other regions and foreign students are more likely to use creative finance tools such as crowdfunding than regions with fewer incoming student flows.

## 5    Discussion and Conclusion

In this paper, we examined the role of knowledge spillover across the Italian regions and their impact in the use of equity crowdfunding campaigns for financing start-ups and projects. We found that student mobility and knowledge flow play a role in the relationships between crowdfunding campaigns and entrepreneurial financial sources in the region under consideration.

Our results show that inbound knowledge in the region, measured in terms of students from other regions, has a significant effect on the adoption of crowdfunding campaigns in the region itself. The wide variety of knowledge from other regions increases the innovation rate [92] and the technological development in the region itself [93, 94], which may stimulate entrepreneurs to use uncommon financial instruments such as equity crowdfunding.

Additionally, we also found that a flow of international students encourages the entrepreneurial mindset in the region itself to use financial sources such as crowdfunding. The diffusion of knowledge within the region is a phenomenon that increases economic development. The presence of students from other regions and other countries has always represented a flow of knowledge that contributes to the development of the region in terms of competitiveness [95] and performance [96]. The use of financial instruments such as equity crowdfunding is an example of how the variety of knowledge and the exchange of knowledge increases the possibility for people to follow and benefit from technological development.

### 5.1 Limitation and Future Research

This work is not without limitations. First, our study is restricted to the context of equity crowdfunding. An investigation of rewards-based crowdfunding, business angels, Initial Coin Offering (ICO), and venture cprevapitals might provide important insights into the role of knowledge spillover in accessing an uncommon and innovative financial source.

Second, we based our research only on Italy. Focusing on one single country ented us from capturing cultural, political, and economic differences in the use of financial sources like crowdfunding campaigns.

Finally, we are not aware of whether the campaign has been successful or not. However, this information should not hamper our findings because we were not looking at the outcome of the campaign but at the impact of knowledge and human capital flow on the adoption of crowdfunding campaigns by users to directly seek financial help from the general public (the "crowd") instead of turning to financial investors such as business angels, banks, or venture capital funds.

### 5.2 Theoretical and Managerial Implications

Our findings have interesting implications for the growing literature on crowdfunding, and more broadly for the entrepreneurship literature. Our findings indicate that a wide variety of inbound knowledge flow can serve as a driver for the development of crowdfunding and other alternative sources of financing.

This relationship can be linked to the natural relationship between entrepreneur and financial market and improve access to the new forms of financial channels. As we can see, this relationship improves start-up development because crowdfunding has the potential to democratize access to capital as it can be a viable option for entrepreneurs who have difficulty accessing traditional channels of financing [8].

These implications are strongly related to regional policies, and our results have several implications for regional development and innovation policies. Regions with heterogeneity in knowledge and human capital are more likely to support the innovation process and the adoption of alternative financial sources such as crowdfunding. Therefore, it is important to understand the role that region attractiveness, in terms of knowledge flow, plays in supporting regional development [12] and what elements contribute to the development of a crowdfunding campaign in a certain area.

Students moving from a region or country are attracted to innovative areas to which they can contribute and from which they can learn, they may effectively be part of a virtuous cycle where a strong innovation system attracts skills, which make it stronger and, in turn, more attractive to human capital [63]. As a result, this heterogeneity of incoming knowledge flow participates in the development of the area itself in terms of innovation and, in this case, in the use of alternative financial sources such as crowdfunding. Several important factors impact the use of crowdfunding campaigns in the local context, such investment [97], entrepreneurial complexity [98], and financial access [16]. Despite its limitations, however, the analysis shows the importance of looking at incoming knowledge flows and calls for further research on the implications of different types of mobility and the resulting impact on crowdfunding development in the local context.

# References

1. Giudici, G., Nava, R., Rossi Lamastra, C., Verecondo, C.: Crowdfunding: the new frontier for financing entrepreneurship? Available at SSRN 2157429 (2012)
2. Mollick, E.: The dynamics of crowdfunding: an exploratory study. J. Bus. Ventur. **29**(1), 1–16 (2014)
3. Ahlers, G.K., Cumming, D., Günther, C., Schweizer, D.: Signaling in equity crowdfunding. Entrep. Theory Pract. **39**(4), 955–980 (2015)
4. Guiso, L., Sapienza, P., Zingales, L.: Does local financial development matter? Q. J. Econ. **119**(3), 929–969 (2004)
5. Alessandrini, P., Presbitero, A.F., Zazzaro, A.: Banks, distances and firms' financing constraints. Rev. Financ. **13**(2), 261–307 (2009)
6. Sun, W., Li, Q., Li, B.: Does geographic distance have a significant impact on enterprise financing costs? J. Geog. Sci. **29**(12), 1965–1980 (2019). https://doi.org/10.1007/s11442-019-1699-6
7. Belleflamme, P., Lambert, T., Schwienbacher, A.: Crowdfunding: tapping the right crowd. J. Bus. Ventur. **29**(5), 585–609 (2014)
8. Kim, K., Hann, I.H.: Does crowdfunding democratize access to capital? A geographical analysis. SSRN Electron. J. 1–35 (2013)
9. Agrawal, A., Catalini, C., Goldfarb, A.: Crowdfunding: geography, social networks, and the timing of investment decisions. J. Econ. Manag. Strategy **24**(2), 253–274 (2015)
10. Giudici, G., Guerini, M., Rossi, C.: Why crowdfunding projects can succeed: the role of proponents' territorial social capital. In SMS Conference, pp. 1–20. ESP (2014)
11. Sorenson, O., Assenova, V., Li, G.C., Boada, J., Fleming, L.: Expand innovation finance via crowdfunding. Science **354**(6319), 1526–1528 (2016)
12. Lee, S.Y., Florida, R., Gates, G.: Innovation, human capital, and creativity. Int. Rev. Public Adm. **14**(3), 13–24 (2010)
13. Kahle, K.M., Stulz, R.M.: Access to capital, investment, and the financial crisis. J. Financ. Econ. **110**(2), 280–299 (2013)
14. Block, J., Sandner, P.: What is the effect of the financial crisis on venture capital financing? Empirical evidence from US Internet start-ups. Ventur. Cap. **11**(4), 295–309 (2009)

15. Kirby, E., Worner, S.: Crowd-funding: an infant industry growing fast. IOSCO Research Department (2014)
16. Langley, P., Leyshon, A.: Platform capitalism: the intermediation and capitalization of digital economic circulation. Financ. Soc. **3**(1), 11–31 (2017)
17. Block, J., Hornuf, L., Moritz, A.: Which updates during an equity crowdfunding campaign increase crowd participation? Small Bus. Econ. **50**(1), 3–27 (2017). https://doi.org/10.1007/s11187-017-9876-4
18. Vismara, S.: Signaling to overcome inefficiencies in crowdfunding markets. In: The Economics of Crowdfunding, pp. 29–56. Palgrave Macmillan, Cham (2018)
19. Bapna, S.: Complementarity of signals in early-stage equity investment decisions: evidence from a randomized field experiment. Manage. Sci. **65**(2), 933–952 (2019)
20. Hornuf, L., Schwienbacher, A.: Market mechanisms and funding dynamics in equity crowdfunding. J. Corp. Finan. **50**, 556–574 (2018)
21. Short, J.C., Ketchen, D.J., Jr., McKenny, A.F., Allison, T.H., Ireland, R.D.: Research on crowdfunding: reviewing the (very recent) past and celebrating the present. Entrep. Theory Pract. **41**(2), 149–160 (2017)
22. Lukkarinen, A., Teich, J.E., Wallenius, H., Wallenius, J.: Success drivers of online equity crowdfunding campaigns. Decis. Support Syst. **87**, 26–38 (2016)
23. Piva, E., Rossi-Lamastra, C.: Human capital signals and entrepreneurs' success in equity crowdfunding. Small Bus. Econ. **51**(3), 667–686 (2017). https://doi.org/10.1007/s11187-017-9950-y
24. Barbi, M., Mattioli, S.: Human capital, investor trust, and equity crowdfunding. Res. Int. Bus. Financ. **49**, 1–12 (2019)
25. Bernstein, S., Korteweg, A., Laws, K.: Attracting early-stage investors: evidence from a randomized field experiment. J. Financ. **72**(2), 509–538 (2017)
26. Bollaert, H., Leboeuf, G., Schwienbacher, A.: The narcissism of crowdfunding entrepreneurs. Small Bus. Econ. **55**(1), 57–76 (2019). https://doi.org/10.1007/s11187-019-00145-w
27. Schultz, T.W.: Investment in human capital: reply. Am. Econ. Rev. **51**(5), 1035–1039 (1961)
28. Becker, G.S.: Investment in human capital: effects on earnings (No. c11230). National Bureau of Economic Research (1994)
29. Becker, G.S.: Investment in human capital: a theoretical analysis. J. Polit. Econ. **70**(5, Part 2), 9–49 (1962)
30. Mariz-Pérez, R.M., Teijeiro-Álvarez, M.M., García-Álvarez, M.T.: The relevance of human capital as a driver for innovation. Cuadernos Econ. **35**(98), 68–76 (2012)
31. Goldin, C.D.: Human capital. In: Handbook of Cliometrics (2016)
32. Ucbasaran, D., Westhead, P., Wright, M.: Opportunity identification and pursuit: does an entrepreneur's human capital matter? Small Bus. Econ. **30**(2), 153–173 (2008)
33. Unger, J.M., Rauch, A., Frese, M., Rosenbusch, N.: Human capital and entrepreneurial success: a meta-analytical review. J. Bus. Ventur. **26**(3), 341–358 (2011)
34. Chatterjee, S., Lubatkin, M.H., Schweiger, D.M., Weber, Y.: Cultural differences and shareholder value in related mergers: linking equity and human capital. Strateg. Manag. J. **13**(5), 319–334 (1992)
35. Frid, C.J.: Acquiring financial resources to form new ventures: the impact of personal characteristics on organizational emergence. J. Small Bus. Entrep. **27**(3), 323–341 (2014)
36. Glaeser, E.: The new economics of urban and regional growth. In: The Oxford Handbook of Economic Geography, pp. 83–99 (2000)
37. Knoben, J., Oerlemans, L.A.G.: Ties that spatially bind? A relational account of the causes of spatial firm mobility. Reg. Stud. **42**(3), 385–400 (2008)
38. Faggian, A., McCann, P.: Human capital flows and regional knowledge assets: a simultaneous equation approach. Oxf. Econ. Pap. **58**(3), 475–500 (2006)

39. Faggian, A., McCann, P.: Human capital, graduate migration and innovation in British regions. Camb. J. Econ. **33**(2), 317–333 (2009)
40. Faggian, A., McCann, P.: Universities, agglomerations and graduate human capital mobility. Tijdschr. Econ. Soc. Geogr. **100**(2), 210–223 (2009)
41. Caragliu, A., Nijkamp, P.: Space and knowledge spillovers in European regions: the impact of different forms of proximity on spatial knowledge diffusion. J. Econ. Geogr. **16**(3), 749–774 (2016)
42. Iammarino, S., McCann, P.: The structure and evolution of industrial clusters: transactions, technology and knowledge spillovers. Res. Policy **35**(7), 1018–1036 (2006)
43. Abramovsky, L., Simpson, H.: Geographic proximity and firm–university innovation linkages: evidence from great britain. J. Econ. Geogr. **11**(6), 949–977 (2011)
44. Capello, R., Faggian, A.: Collective learning and relational capital in local innovation processes. Reg. Stud. **39**(1), 75–87 (2005)
45. Romer, P.M.: Endogenous technological change. J. Polit. Econ. **98**(5, Part 2), S71–S102 (1990)
46. Romer, P.M.: Increasing returns and long-run growth. J. Polit. Econ. **94**(5), 1002–1037 (1986)
47. Romer, P.M.: The origins of endogenous growth. J. Econ. Perspect. **8**(1), 3–22 (1994)
48. Lucas, R.E., Jr.: On the mechanics of economic development. J. Monet. Econ. **22**(1), 3–42 (1988)
49. Dotti, N.F., Fratesi, U., Lenzi, C., Percoco, M.: Local labour markets and the interregional mobility of Italian university students. Spat. Econ. Anal. **8**(4), 443–468 (2013)
50. Giambona, F., Porcu, M., Sulis, I.: Students mobility: assessing the determinants of attractiveness across competing territorial areas. Soc. Indic. Res. **133**(3), 1105–1132 (2017)
51. Florida, R.: Toward the learning region. Futures **27**(5), 527–536 (1995)
52. Sedlacek, S.: The role of universities in fostering sustainable development at the regional level. J. Clean. Prod. **48**, 74–84 (2013)
53. Coe, N.M., Hess, M., Yeung, H.W.C., Dicken, P., Henderson, J.: 'Globalizing' regional development: a global production networks perspective. Trans. Inst. Br. Geogr. **29**(4), 468–484 (2004)
54. Goldstein, H., Drucker, J.: The economic development impacts of universities on regions: do size and distance matter? Econ. Dev. Q. **20**(1), 22–43 (2006)
55. King, R., Ruiz-Gelices, E.: International student migration and the European 'year abroad': effects on European identity and subsequent migration behaviour. Int. J. Popul. Geogr. **9**(3), 229–252 (2003)
56. Dustmann, C., Glitz, A.: Migration and education. In: Handbook of the Economics of Education, vol. 4, pp. 327–439. Elsevier (2011)
57. Fini, R., Grimaldi, R., Meoli, A.: The effectiveness of university regulations to foster science-based entrepreneurship. Res. Policy **49**(10), 104048 (2020)
58. Townsend, D.M., Hunt, R.A., McMullen, J.S., Sarasvathy, S.D.: Uncertainty, knowledge problems, and entrepreneurial action. Acad. Manag. Ann. **12**(2), 659–687 (2018)
59. Maurseth, P.B., Verspagen, B.: Knowledge spillovers in Europe: a patent citations analysis. Scand. J. Econ. **104**(4), 531–545 (2002)
60. Harabi, N.: Channels of R&D spillovers: an empirical investigation of Swiss firms. Technovation **17**(11–12), 627–635 (1997)
61. Florida, R., Kenney, M.: Restructuring in place: Japanese investment, production organization, and the geography of steel. Econ. Geogr. **68**(2), 146–173 (1992)
62. Florida, R., Kenney, M.: Venture capital and high technology entrepreneurship. J. Bus. Ventur. **3**(4), 301–319 (1988)
63. Marinelli, E.: Sub-national graduate mobility and knowledge flows: an exploratory analysis of onward-and return-migrants in Italy. Reg. Stud. **47**(10), 1618–1633 (2013)

64. Haber, S., Reichel, A.: The cumulative nature of the entrepreneurial process: the contribution of human capital, planning and environment resources to small venture performance. J. Bus. Ventur. **22**(1), 119–145 (2007)
65. Zhou, Y., Fan, X., Son, J.: How and when matter: exploring the interaction effects of high-performance work systems, employee participation, and human capital on organizational innovation. Hum. Resour. Manage. **58**(3), 253–268 (2019)
66. El Shoubaki, A., Laguir, I., den Besten, M.: Human capital and SME growth: the mediating role of reasons to start a business. Small Bus. Econ. 1–15 (2019)
67. Eriksson, R., Rataj, M.: The geography of starts-ups in Sweden. The role of human capital, social capital and agglomeration. Entrep. Reg. Dev. **31**(9–10), 735–754 (2019)
68. Easley, D., O'hara, M.: Information and the cost of capital. J. Financ. **59**(4), 1553–1583 (2004)
69. Chen, H., Gompers, P., Kovner, A., Lerner, J.: Buy local? The geography of venture capital. J. Urban Econ. **67**(1), 90–102 (2010)
70. Cumming, D., Dai, N.: Local bias in venture capital investments. J. Empir. Financ. **17**(3), 362–380 (2010)
71. Nanda, R., Rhodes-Kropf, M.: Investment cycles and startup innovation. J. Financ. Econ. **110**(2), 403–418 (2013)
72. Green, K.M., Covin, J.G., Slevin, D.P.: Exploring the relationship between strategic reactiveness and entrepreneurial orientation: the role of structure–style fit. J. Bus. Ventur. **23**(3), 356–383 (2008)
73. Dicken, P., Malmberg, A.: Firms in territories: a relational perspective. Econ. Geogr. **77**(4), 345–363 (2001)
74. Rothgang, M., et al.: Cluster policy: insights from the German leading edge cluster competition. J. Open Innov.: Technol. Mark. Complex. **3**(1), 1–20 (2017). https://doi.org/10.1186/s40852-017-0064-1
75. Porter, M.E.: Clusters and the new economics of competition, vol. 76, no. 6, pp. 77–90. Harvard Business Review, Boston (1998)
76. Sorenson, O., Rivkin, J.W., Fleming, L.: Complexity, networks and knowledge flow. Res. Policy **35**(7), 994–1017 (2006)
77. Rocha, H.O., Sternberg, R.: Entrepreneurship: the role of clusters theoretical perspectives and empirical evidence from Germany. Small Bus. Econ. **24**(3), 267–292 (2005)
78. Josefy, M., Dean, T.J., Albert, L.S., Fitza, M.A.: The role of community in crowdfunding success: evidence on cultural attributes in funding campaigns to "save the local theater." Entrep. Theory Pract. **41**(2), 161–182 (2017)
79. Giudici, G., Guerini, M., Rossi-Lamastra, C.: Reward-based crowdfunding of entrepreneurial projects: the effect of local altruism and localized social capital on proponents' success. Small Bus. Econ. **50**(2), 307–324 (2017). https://doi.org/10.1007/s11187-016-9830-x
80. Colombo, M.G., Franzoni, C., Rossi-Lamastra, C.: Internal social capital and the attraction of early contributions in crowdfunding. Entrep. Theory Pract. **39**(1), 75–100 (2015)
81. Paravisini, D.: Local bank financial constraints and firm access to external finance. J. Financ. **63**(5), 2161–2193 (2008)
82. Tödtling, F., Kaufmann, A.: The role of the region for innovation activities of SMEs. Eur. Urban Reg. Stud. **8**(3), 203–215 (2001)
83. Freeman, R.B.: Globalization of scientific and engineering talent: international mobility of students, workers, and ideas and the world economy. Econ. Innov. New Technol. **19**(5), 393–406 (2010)
84. Trippl, M.: Scientific mobility and knowledge transfer at the interregional and intraregional level. Reg. Stud. **47**(10), 1653–1667 (2013)
85. Anselin, L., Varga, A., Acs, Z.: Geographical spillovers and university research: a spatial econometric perspective. Growth Chang. **31**(4), 501–515 (2000)

86. Park, G., Shin, S.R., Choy, M.: Early mover (dis) advantages and knowledge spillover effects on blockchain startups' funding and innovation performance. J. Bus. Res. **109**, 64–75 (2020)
87. González, C.R., Mesanza, R.B., Mariel, P.: The determinants of international student mobility flows: an empirical study on the Erasmus programme. High. Educ. **62**(4), 413–430 (2011)
88. Guruz, K.: Higher Education and International Student Mobility in the Global Knowledge Economy: Revised and Updated Second Edition. Suny Press (2011)
89. Choudaha, R., Chang, L.: Trends in international student mobility. World Educ. News Rev. **25**(2) (2012)
90. Siekierski, P., Lima, M.C., Borini, F.M., Pereira, R.M.: International academic mobility and innovation: a literature review. J. Glob. Mob. Home Expatriate Manage. Res. (2018)
91. Beyhan, B., Findik, D.: Student and graduate entrepreneurship: ambidextrous universities create more nascent entrepreneurs. J. Technol. Transf. **43**(5), 1346–1374 (2017). https://doi.org/10.1007/s10961-017-9590-z
92. Fallah, M.H., Ibrahim, S.: Knowledge spillover and innovation in technological clusters. In: Proceedings of IAMOT 2004 Conference, Washington, DC, pp. 1–16, April 2004
93. Huffman, D., Quigley, J.M.: The role of the university in attracting high tech entrepreneurship: a Silicon Valley tale. Ann. Reg. Sci. **36**(3), 403–419 (2002)
94. Nieto, M., Quevedo, P.: Absorptive capacity, technological opportunity, knowledge spillovers, and innovative effort. Technovation **25**(10), 1141–1157 (2005)
95. Audretsch, D.B., Bönte, W., Keilbach, M.: Entrepreneurship capital and its impact on knowledge diffusion and economic performance. J. Bus. Ventur. **23**(6), 687–698 (2008)
96. Gilbert, B.A., McDougall, P.P., Audretsch, D.B.: Clusters, knowledge spillovers and new venture performance: an empirical examination. J. Bus. Ventur. **23**(4), 405–422 (2008)
97. Yu, S., Johnson, S., Lai, C., Cricelli, A., Fleming, L.: Crowdfunding and regional entrepreneurial investment: an application of the CrowdBerkeley database. Res. Policy **46**(10), 1723–1737 (2017)
98. Audretsch, D.B., Hülsbeck, M., Lehmann, E.E.: Regional competitiveness, university spillovers, and entrepreneurial activity. Small Bus. Econ. **39**(3), 587–601 (2012)